\documentclass[conference]{IEEEtran}

\IEEEoverridecommandlockouts

\usepackage[utf8]{inputenc}

\usepackage{csquotes}

\usepackage[numbers]{natbib}

\usepackage{graphicx}
\graphicspath{{images/}}

\usepackage{amsmath}

\usepackage{eurosym}

\PassOptionsToPackage{hyphens}{url}\usepackage{hyperref}

\usepackage{array}

\usepackage{multirow}

\usepackage{makecell}

\usepackage{float}
\usepackage{dblfloatfix}

\usepackage{tcolorbox}
\definecolor{amber}{rgb}{1.0, 0.75, 0.0}
\newtcolorbox{mybox}{colback=amber!10,colframe=amber}

\begin{document}

\title{
    How Do Microservice API Patterns Impact Understandability? A Controlled Experiment{}\thanks{\IEEEauthorrefmark{1}Research participation while at University of Stuttgart, Germany}
}

\author{
    \IEEEauthorblockN{Justus Bogner}
    \IEEEauthorblockA{
        Vrije Universiteit Amsterdam\\
        Amsterdam, The Netherlands\\
        j.bogner@vu.nl
    }
    \and
    \IEEEauthorblockN{Pawel Wójcik}
    \IEEEauthorblockA{
        Independent researcher\IEEEauthorrefmark{1}\\
        Stuttgart, Germany\\
        pawel.wojcik@gmx.net
    }
    \and
    \IEEEauthorblockN{Olaf Zimmermann}
    \IEEEauthorblockA{
        OST Eastern Switzerland University of Applied Sciences\\
        Rapperswil, Switzerland\\
        olaf.zimmermann@ost.ch
    }
}

\maketitle


\begin{abstract}
Microservices expose their functionality via remote Application Programming Interfaces (APIs), e.g., based on HTTP or asynchronous messaging technology.
To solve recurring problems in this design space, Microservice API Patterns (MAPs) have emerged to capture the collective experience of the API design community.
At present, there is a lack of empirical evidence for the effectiveness of these patterns, e.g., how they impact understandability and API usability.

We therefore conducted a controlled experiment with 6 microservice patterns to evaluate their impact on understandability with 65 diverse participants.
Additionally, we wanted to study how demographics like years of professional experience or experience with MAPs influence the effects of the patterns.
Per pattern, we constructed two API examples, each in a pattern version \texttt{P} and a functionally equivalent non-pattern version \texttt{N} (24 in total).
Based on a crossover design, participants had to answer comprehension questions, while we measured the time.

For five of the six patterns, we identified a significant positive impact on understandability, i.e., participants answered faster and / or more correctly for \texttt{P}.
However, effect sizes were mostly small, with one pattern showing a medium effect.
The correlations between performance and demographics seem to suggest that certain patterns may introduce additional complexity;
people experienced with MAPs will profit more from their effects.
This has important implications for training and education around MAPs and other patterns.
\end{abstract}


\section{Introduction}
Microservices promote an architectural style that focuses on a small amount of business-relevant functionality per service, independent deployment, and lightweight communication mechanisms~\cite{Newman2015}.
Individual services expose their functionality via clearly defined Application Programming Interfaces (APIs)~\cite{Jacobson2011}, often based on RESTful HTTP~\cite{Richardson2007}, but also gRPC and GraphQL.
These service APIs are then accessed by client applications, automation scripts, or other services during service composition.

API design has a strong influence on software quality.
It is therefore important that software engineers using or integrating APIs are able to  understand and interact with them efficiently.
Bad API design decreases API usability~\cite{Myers2016}, thereby leading to increased integration efforts or even misunderstandings that result in incorrect API usage.
Over the years, several papers have synthesized \textit{Microservice API Patterns} (MAPs)~\cite{Zimmermann2017,Stocker2018,Lubke2019,Zimmermann2020}, providing reusable blueprints for solving recurring problems around microservice API design, with the goal to improve different software quality aspects.
More recently, these patterns have been holistically described in a book about MAPs~\cite{MainBook}.
\citet{SecondaryBook} also described an assortment of patterns for microservices in book form, focusing on infrastructure design.

While software design patterns are intended to increase software quality, we currently lack empirical evidence for the recently synthesized MAPs.
While some MAPs have been studied from the perspective of performance efficiency and reliability~\cite{ElMalki2023}, we have insufficient knowledge of how and under which circumstances they improve quality attributes like understandability or API usability, neglecting the perspective of human API consumers.
This is especially important because secondary studies on different types of patterns have shown that their positive impact is by no means guaranteed~\cite{Related:LiteratureSurvey,Related:Wedyan}.
In the face of the popularity of microservices and APIs in general, it becomes crucial to start the accumulation of such evidence for MAPs, thereby helping to inform practitioner's decision-making around API design and pattern selection.

To address this gap, we present a controlled experiment~\cite{Wohlin2012} with 65 participants about the influence of 6 MAPs on understandability and API usability.
Participants had to answer comprehension questions about API examples that either contained a MAP or not, while the time to answer was recorded.
We also collected demographic attributes to study how these influenced the effectiveness of the patterns.
To the best of our knowledge, this is the first empirical study about the impact of MAPs on these quality attributes.

\section{Background and Related Work}

\subsection{Microservice API Patterns (MAPs)}
Originating in civil architecture~\cite{Alexander1977}, patterns are reusable blueprints that support practitioners in solving commonly occurring design problems~\cite{Gamma1994}.
The MAP language captures proven solutions to problems commonly encountered when specifying, implementing, and maintaining message-based APIs~\cite{MainBook}.
MAPs focus on message representations – the payloads exchanged when APIs are called.
These payloads vary in their structure.
The chosen representation structure strongly influences the design time and runtime qualities of an API.
Additionally, the evolution of API specifications and their implementations has to be governed. 
To address these challenges, the MAP language is organized into five categories~\cite{MainBook}: 
\textit{Foundation Patterns}, \textit{Responsibility Patterns}, \textit{Structure Patterns}, \textit{Quality Patterns}, and \textit{Evolution Patterns}.
Many complementary pattern languages for the integration domain exist, for which we refer to \citet{Zimmermann2017}.

Remote APIs are important components in distributed systems; hence, many of the pattern forces, i.e., the design concerns that make the solved problem challenging, qualify as architecturally significant requirements~\cite{Chen2013}.
For instance, sending few but large messages causes less but more intense preparation and processing in the API client and API endpoint; conversely, many small messages are easier to prepare and process, but cause more traffic in the communication infrastructure, e.g., via networking and message serialization.
These API design alternatives can differ not only regarding their performance, e.g., latency, but also regarding qualities such as usability and maintainability.
Each design decision usually impacts some qualities positively and others negatively; hence, design patterns are concerned with trade-offs~\cite{Barney2012}.

\subsection{Related Work}
A large body of research has investigated the quality impact of general design patterns.
For example, \citet{Related:Wedyan} conducted a systematic literature review with 50 primary studies published between 2000 and 2018 that analyzed the impact of the original \enquote{Gang of Four} design patterns~\cite{Gamma1994}.
The authors state that many existing studies have different objectives, quality metrics, or attributes and thus lead to conflicting evidence for the effectiveness of patterns to improve quality.
Furthermore, they found that attributes like the size of pattern classes or the documentation of patterns also impacted the quality of the examined systems.

Some studies analyzed API quality without patterns.
\citet{Bogner2023} examined the effect of design rules for RESTful APIs on understandability via a controlled experiment with 105 participants.
For each rule and a complementary violation, participants answered comprehension questions based on Web API snippets and also rated their perceived difficulty.
Most rules had a significant positive impact on understandability, both for the comprehension tasks and difficulty ratings.
Another example is by \citet{Fletcher:2019}, who proposed a quality-aware Web API recommendation system for mashup development.
The approach uses matrix factorization to recommend high-quality web APIs to use.
Their API quality model is based on functionality, reliability, and usability, with understandability only being considered through the existence of different types of documentation.

Several publications investigated patterns in the realm of service orientation.
\citet{Related:Bogner} examined four service-based design patterns regarding their impact on evolvability through a metric-based analysis and controlled experiment with 69 BSc students, who had to extend two versions of a service-based system (one with patterns, one without them).
Although mean efficiency was 12\% higher in the treatment group, this difference was only statistically significant for one of the three tasks.
The authors concluded that there is no clear empirical evidence for evolvability improvements.

\citet{Related:Vale} organized semi-structured interviews with nine industry experts to identify the software quality trade-offs when using design patterns for microservices.
While the interviews confirmed some of the trade-offs already included in the pattern documentations, several new drawbacks could be derived.
According to the interviewees, three patterns displayed more negative than positive characteristics, two showed an equal amount of positive and negative points, and nine patterns had substantially more gains than drawbacks.

One quality attribute that has been studied more frequently in relation to microservice patterns is performance efficiency.
For example, \citet{Pinciroli:2023} studied patterns like \textit{Anti-Corruption Layer}, \textit{Backends for Frontends}, and \textit{CQRS} using performance models based on queuing networks.
They concluded that most patterns conform to practitioners' positive expectations.
Using benchmarking instead of simulation, \citet{Related:Akbulut} analyzed the impact of the three patterns \textit{API Gateway}, \textit{Chain of Responsibility}, and \textit{Asynchronous Messaging}.
The authors emphasized that none of the three patterns significantly outperformed the others in general.
Rather, each pattern excelled in a specific usage scenario.
El Malki et al. even evaluated two MAPs, namely \textit{Request Bundle} from a performance perspective~\cite{ElMalki2021} and \textit{Rate Limit} from a reliability perspective~\cite{ElMalki2022}, with a positive impact in both cases.

In a slightly different MAP-related study, \citet{Bakhtin2022} performed a gray literature review to find and catalog available tools to detect MAPs in software systems.
They also investigated the concrete mechanisms to detect these patterns, and concluded that the identification of 34 of the 46 MAPs they studied is supported by existing tools.

In summary, few empirical studies on the quality impact of microservice patterns exist, and even less so for MAPs.
The few existing studies for MAPs focused on performance efficiency and reliability, leaving open the human perspective of an API consumer, also known as developer experience.

\section{Study Design}
In this section, we describe our detailed experiment design.
Following \citet{Wyrich2023}, we provide the most important study characteristics in Table~\ref{tab:experimentOverview} for a quick overview.

\begin{table}[ht]
    \caption{Experiment overview}
    \label{tab:experimentOverview}
    \centering
    \begin{tabular}{l p{0.65\columnwidth}}
        \hline
        \hline
        Goal & Study the impact of MAPs on understandability\\
        Study objects & 6 MAPs described by \citet{MainBook} and \citet{SecondaryBook}, 2 API examples per MAP, each example in 2 equivalent versions (one with a pattern, one without it)\\
        Participants & 65 people with at least basic software engineering experience (28 students, 37 professionals from both academia and industry)\\
        Setting & Online experiment via LimeSurvey\\
        Dependent variables & Timed Actual Understandability (TAU)~\cite{DefinitionTAU}, an aggregation of correctness and time\\
        Treatments & API example with a pattern (version \texttt{P}) or without a pattern (version \texttt{N})\\
        \makecell[lt]{Other independent\\variables} & Demographic participant attributes like pattern experience or current role\\
        Tasks & Answering comprehension questions about API examples (12 per participant, 24 in total)\\
        Design & Crossover design with 2 counterbalanced sequences, each with 6 tasks for \texttt{P} and 6 for \texttt{N}\\
        \hline
        \hline
    \end{tabular}
\end{table}

\subsection{Research Questions}
We investigated two research questions, RQ1 and RQ2.

\textbf{RQ1:} How do Microservice API Patterns (MAPs) impact the understandability of service interfaces?

This RQ represents the main focus of our study: we want to provide empirical evidence for the effectiveness of MAPs to improve software quality, in this case the \textit{understandability} of service interfaces from the perspective of human API consumers.
In some cases, the experiment task goes a bit beyond understandability and touches \textit{API usability}~\cite{Myers2016}, a more comprehensive quality attribute describing how effectively and efficiently developers can interact with an API, which is sometimes also referred to as \enquote{DevX}.
All patterns introduce some additional complexity, which needs to be compared against the problem that the pattern solves.

\textbf{RQ2:} How does the demographic background of API consumers influence the effectiveness of the patterns?

Our second RQ is of an exploratory nature.
We want to analyze if and how demographic characteristics of API consumers influence the effect that the patterns have on understandability.
For example, it might be possible that MAPs only show a reduced effect on students when compared to professionals.
Some patterns might also be so complex that their effect only manifests with participants having solid experience with microservices and MAPs.
We selected a few general attributes plus several specific ones for the experiment context during the study design phase.

\subsection{Participants and Sampling}
Considering our second RQ, our goal was to attract participants from diverse backgrounds.
We therefore set the minimum requirements for participation fairly low.
Participants needed to understand English and should have at least basic software engineering expertise.
To ensure this, we accepted MSc and PhD students from study programs related to computer science, but no BSc students, who might not have enough software engineering expertise yet.
Additionally, we accepted software or IT professionals working in industry or academia.

To recruit participants, we relied on \textit{convenience sampling} together with \textit{referral-chain sampling}~\cite{Baltes2022}.
We shared the call for participation within our personal networks via email, and asked participants to further distribute it within their network.
We also shared it with students via mailing lists of the university.
Furthermore, the call was distributed via several social media networks, like LinkedIn\footnote{\url{https://www.linkedin.com}} and Twitter\footnote{\url{https://www.twitter.com}}.
It was also shared in company-internal Microsoft Teams Channels related to microservices.
To encourage participation, we also pledged the donation of 1~\EUR{} for each of the first 200 fully completed responses to charity.

\subsection{Experiment Objects and Tasks}
Our experiment objects were concrete MAPs, which meant that there were many patterns from different categories to choose from.
Since MAPs can impact software quality in different ways and also focus on different quality aspects, we analyzed the pattern collections regarding patterns for which we expected an impact on understandability and API usability.
Furthermore, we analyzed for which patterns we could easily create tasks for an online experiment.
In the end, we selected six patterns, which are summarized in Table~\ref{tab:selectedPatterns} and described with their problem-solution pairs plus an ID below.
Strictly speaking, \textit{API Gateway} is not a pattern from the MAP book. As it still is a very relevant pattern in the space of microservice APIs, we decided to include it.
For more information about the patterns, please refer to their sources.

\begin{table}[H]
    \caption{The six patterns selected for the experiment}
    \label{tab:selectedPatterns}
    \centering
    \begin{tabular}{l l}
        \textbf{Pattern Name} & \textbf{Source}\\
        \hline
        \hline
        Error Report & \citet{MainBook}\\
        Parameter Tree & \citet{MainBook}\\
        Wish List & \citet{MainBook}\\
        Metadata Element & \citet{MainBook}\\
        Request Bundle & \citet{MainBook}\\
        API Gateway & \citet{SecondaryBook}\\
        \hline
        \hline
    \end{tabular}
\end{table}

\begin{figure*}[ht]
  \centering
  \includegraphics[width=0.95\textwidth]{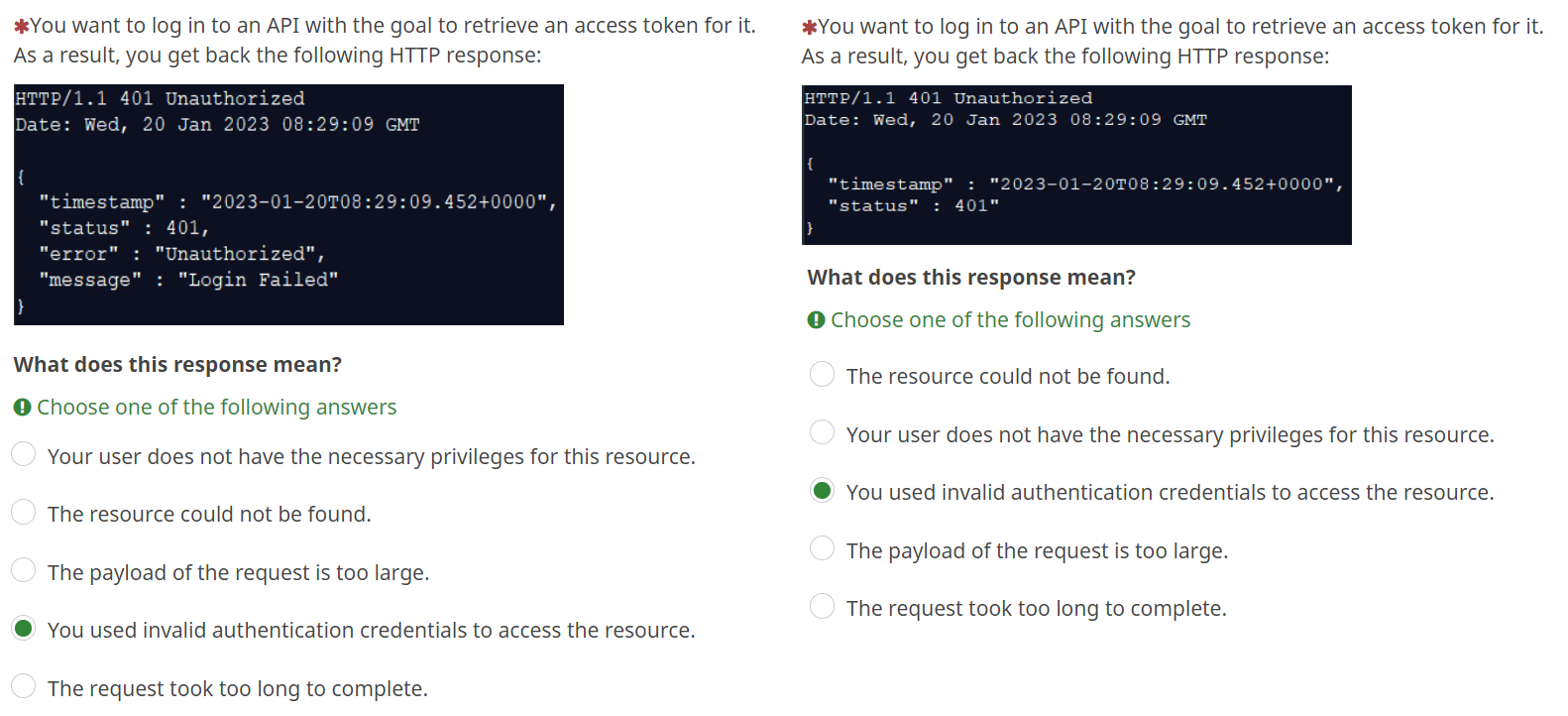}
  \vspace{-3pt}
  \caption{\textit{Error Report 1} as an example of a single choice task (pattern version \texttt{P} on the left, non-pattern version \texttt{N} on the right, correct answers checked)}
  \label{fig:SingleChoiceTask}
\end{figure*}

\textbf{Error Report (ER)}
\begin{itemize}
    \item \textit{Problem:} How can an API provider inform its clients about communication and processing faults? How can this information be made independent of the underlying communication technologies and platforms?
    \item \textit{Solution:} Reply with error codes in response messages that indicate and classify the faults in a simple, machine-readable way. Additionally, add textual descriptions of the errors for the API client stakeholders, including developers and / or end users such as administrators.
\end{itemize}

\textbf{Parameter Tree (PT)}
\begin{itemize}
    \item \textit{Problem:} How can containment relationships be expressed when defining complex representation elements and exchanging such related elements at runtime?
    \item \textit{Solution:} Define a \textit{Parameter Tree}, a hierarchical structure with one root node and one or more child nodes.
\end{itemize}

\textbf{Wish List (WL)}
\begin{itemize}
    \item \textit{Problem:} How can an API client inform the API provider at runtime about the data it is interested in?
    \item \textit{Solution:} As an API client, provide a \textit{Wish List} in the request that enumerates all desired data elements of the requested resource. As an API provider, deliver only those data elements in the response message that are enumerated in the \textit{Wish List}.
\end{itemize}

\textbf{Metadata Element (ME)}
\begin{itemize}
    \item \textit{Problem:} How can messages be enriched with additional information so that receivers can interpret the message content correctly, without having to hard-code assumptions about the data semantics?
    \item \textit{Solution:} Introduce one or more \textit{Metadata Elements} to explain and enhance the other representation elements that appear in request and response messages. Populate the values of these elements thoroughly and consistently. Process them to steer interoperable and efficient message consumption and processing.
\end{itemize}

\textbf{Request Bundle (RB)}
\begin{itemize}
    \item \textit{Problem:} How can the number of requests and responses be reduced to increase communication efficiency?
    \item \textit{Solution:} Define a \textit{Request Bundle} as a data container that assembles multiple independent requests in a single request message. Add metadata such as identifiers of individual requests and bundle element counter.
\end{itemize}

\textbf{API Gateway (AG)}
\begin{itemize}
    \item \textit{Problem:} How do the clients of a microservice-based application access the individual services?
    \item \textit{Solution:} Implement an API Gateway that is the single entry point for all clients. It handles requests in one of two ways. Some requests are simply routed to the appropriate service. Other requests are sent out to multiple services.
\end{itemize}

\begin{figure*}[bt]
  \centering
  \includegraphics[width=0.95\textwidth]{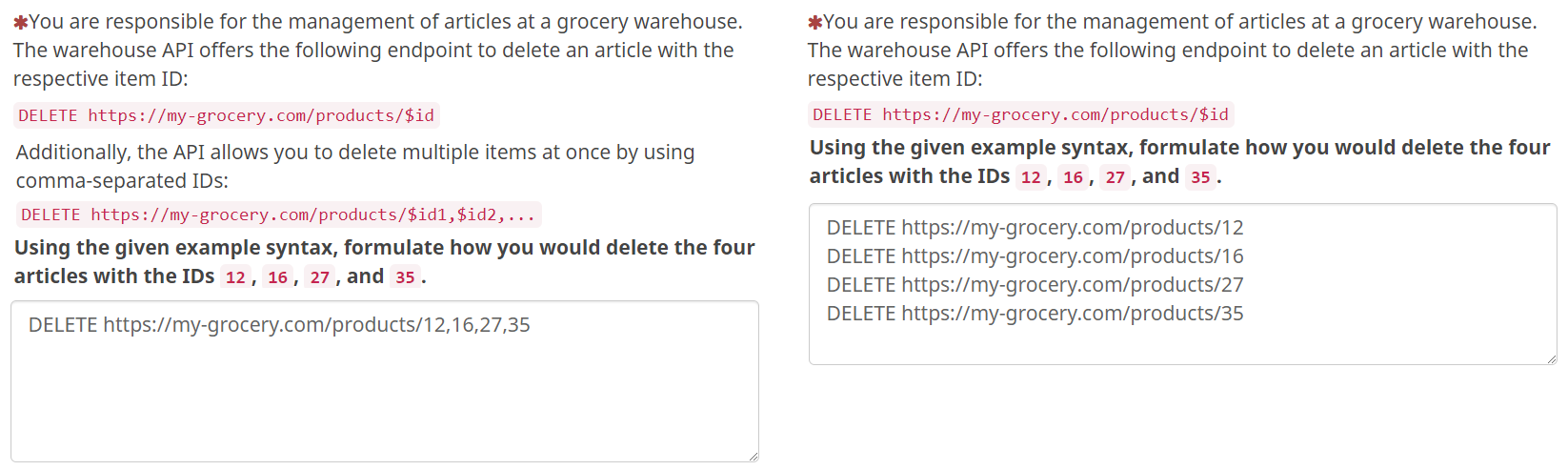}
  \vspace{-3pt}
  \caption{\textit{Request Bundle 2} as an example of a free-text task (pattern version \texttt{P} on the left, non-pattern version \texttt{N} on the right, correct answers filled in)}
  \label{fig:FreeTextTask}
\end{figure*}

For each of these six patterns, we created two different comprehension tasks with two versions each, one containing the pattern (\texttt{P}) and one without the pattern (\texttt{N}).
Each task first presented a concrete HTTP API example, e.g., through a mix of text and figures, and then posed a comprehension question about it.
These questions were mostly single choice questions where participants had to pick the correct answer from five options displayed in random order.
While single choice questions are easy to operationalize and evaluate, there is also a probability that participants randomly guess the correct answer or can derive it by excluding wrong ones, both of which might not convey the full extent of understanding that we are interested in.
We therefore also used some tasks based on free-text questions, where participants had to write the answer themselves, e.g., the names of several entities or a number.
For some of these questions, an expected syntax was provided with the description, e.g., 
for creating an HTTP request.
An example of the two complementary versions of a single choice task can be seen in Fig.~\ref{fig:SingleChoiceTask}, while Fig.~\ref{fig:FreeTextTask} provides the same for a free-text task.
Please refer to our digital appendix for the complete description of all 12 tasks.\footnote{\url{https://doi.org/10.5281/zenodo.10210246}}

\subsection{Experiment Variables and Hypotheses}
The \textbf{dependent variable} per task in this controlled experiment was \textit{Timed Actual Understandability} (TAU), which we adapted from \citet{DefinitionTAU}.
It is an aggregation of \textit{correctness} of answers and the required \textit{time} to answer in seconds, with values between \texttt{0} and \texttt{1}.
In this experiment, TAU for a given participant $p$ and task $t$ with one of the two versions $v$ (either \texttt{P} or \texttt{N}) was calculated as follows: 

\begin{align*}
TAU_{p,t,v} = correctness_{p,t,v} \times ( 1 - \frac{time_{p,t,v}}{max(time_{t})})
\end{align*}

\textit{Correctness} is binary, i.e., either \texttt{1} for a correct answer or \texttt{0} otherwise.
The result is then multiplied with the inverted relation of the required time to the longest time for this task over all participants, including both the \texttt{P} and \texttt{N} versions.
If a participant answered incorrectly, the value of TAU becomes \textit{0}, no matter how fast the response.
The closer TAU is to \texttt{1}, the higher the understandability in general.
The \textbf{independent variables} in this experiment were the systematically controlled \textit{task versions} (\texttt{P} or \texttt{N}) and the \textit{demographic attributes} of the participants.
The concrete attributes collected in this experiment were the \textit{years of professional experience}, \textit{professional background} (industry, academia, both), the main \textit{professional role}, and the self-assessed \textit{experience levels} in the areas of HTTP APIs, general software design patterns, and MAPs.

For RQ1, we expected that the comprehension performance for the pattern version \texttt{P} of a task would be significantly better than for the non-pattern version \texttt{N}.
This can be formalized with the following \textbf{hypothesis} pair:

\textit{Null Hypothesis} (H$_{0}$): 
Task version \texttt{P} leads to lower or equal comprehension performance (TAU) than the complementary task version \texttt{N}.

\textit{Alternative Hypothesis} (H$_{1}$): 
Task version \texttt{P} leads to higher comprehension performance (TAU) than the complementary task version \texttt{N}.

While only this one hypothesis is listed explicitly for RQ1, statistical tests were performed individually per pattern, with the combined two tasks per pattern.
Therefore, six statistical tests were conducted in total to clearly identify which patterns have a significant impact.
For RQ2, we had no clear hypotheses, since it was exploratory in nature.

\subsection{Experiment Design and Execution}
To be more robust towards inter-participant differences and to increase the number of observations per pattern, we opted for a \textit{crossover design}~\cite{Crossover}, a special version of a within-subject design, where participants work on all treatments, but in different orders.
We used the 12 task pairs (2 per pattern) to carefully construct 2 sequences that contained either the \texttt{P} or \texttt{N} version of a task (see Table~\ref{tab:experimentSequences}).
Participants were randomly assigned to one of the sequences.
The goal was to not show longer tasks right after another and to maximize the distance between the versions \texttt{P} and \texttt{N} of the same pattern.
For example, sequence 1 sees the pattern version \texttt{P} of \textit{Error Report 1} (ER1P) as the first task and the non-pattern version \texttt{N} of \textit{Error Report 2} (ER2N) as task 7.
Conversely, sequence 2 sees ER1N as task 1 and ER2P as task 7.

\begin{table*}[bt]
    \caption{Demographics per experiment sequence (mean used for experience, last three columns based on ordinal scale from 1 to 5)}
    \label{tab:demographics}
    \centering
    \begin{tabular}{rrrrrrrr}
        Sequence & \# Participants & \# Students & \# Industry & Years of Exp. & Pattern Exp. & HTTP API Exp. & MAP Exp.\\
        \hline
        \hline
        1 & 35 & 17 & 8 & 7.34 & 3.23 & 3.03 & 2.63 \\
        2 & 30 & 11 & 17 & 5.73 & 2.77 & 2.77 & 2.03 \\
        \hline
        \hline
    \end{tabular}
\end{table*}

Before the actual tasks, an introductory page was shown that prepared the participants for the experiment.
It explained, e.g., the types of tasks to expect, that time would be measured but that it would be more important to answer correctly, that they could withdraw at any time, and what data we collected and what we would do with it.
Participants had to give consent to these terms before being able to start the experiment.
After completing all 12 tasks, participants finally answered the demographic questions.
To make participation as simple as possible, we opted for a Web-based online experiment via a self-hosted \textit{LimeSurvey}\footnote{\url{https://www.limesurvey.org}} instance.
This tool makes it easy to customize the structure and the content of tasks, measures the time per question, and can assign participants randomly to different sequences.
Participants used LimeSurvey via the browser of their choice.

\begin{table}[H]
    \caption{Sequences of pattern (P) and non-pattern (N) versions}
    \label{tab:experimentSequences}
    \centering
    \begin{tabular}{r l l}
        \textbf{\#} & \textbf{Sequence 1} & \textbf{Sequence 2}\\
        \hline
        \hline
        -- & \multicolumn{2}{c}{Experiment Introduction}\\
        1 & ErrorReport1 \texttt{P} (ER1P) & ErrorReport1 \texttt{N} (ER1N)\\
        2 & ParameterTree1 \texttt{N} (PT1N) & ParameterTree1 \texttt{P} (PT1P)\\
        3 & WishList1 \texttt{P} (WL1P) & WishList1 \texttt{N} (WL1N)\\
        4 & MetadataElement1 \texttt{N} (ME1N) & MetadataElement1 \texttt{P} (ME1P)\\
        5 & APIGateway1 \texttt{P} (AG1P) & APIGateway1 \texttt{N} (AG1N)\\
        6 & RequestBundle1 \texttt{N} (RB1N) & RequestBundle1 \texttt{P} (RB1P)\\
        7 & ErrorReport2 \texttt{N} (ER2N) & ErrorReport2 \texttt{P} (ER2P)\\
        8 & ParameterTree2 \texttt{P} (PT2P) & ParameterTree2 \texttt{N} (PT2N)\\
        9 & WishList2 \texttt{N} (WL2N) & WishList2 \texttt{P} (WL2P)\\
        10 & MetadataElement2 \texttt{P} (ME2P) & MetadataElement2 \texttt{N} (ME2N)\\
        11 & APIGateway2 \texttt{N} (AG2N) & APIGateway2 \texttt{P} (AG2P)\\
        12 & RequestBundle2 \texttt{P} (RB2P) & RequestBundle2 \texttt{N} (RB2N)\\
        -- & \multicolumn{2}{c}{Demographic Questions}\\
        \hline
        \hline
    \end{tabular}
\end{table}

This experiment design was the result of several iterations and discussions, with tasks being continuously refined.
We also conducted a pilot experiment with an earlier version and a closed group of six experienced professionals from both academia and industry, one of them even an author of the MAP book~\cite{MainBook}.
This group was asked to provide general feedback about the structure, visual presentation, and task design.
They were also specifically asked to assess the degree of realism of the used API examples and the fairness of comparing the complementary versions.
Most of the feedback was positive, with some minor suggestions for improvement, e.g., using slightly different phrasing or optimizing colors and font size.
However, some examples were criticized as too artificial.
Because of this, we dropped one pattern entirely (we previously had seven) and adjusted several API examples.

\subsection{Experiment Analysis}
In total, 98 participants started the experiment and 66 completed it.
Only complete responses were used for the analysis, with the remaining 32 being discarded.
The majority of these belonged to participants that stopped the survey within the first two questions.
Reasons might be the realization of the size or complexity of the tasks, or simple curiosity that tempted users to just view the first one or two tasks.
After a two-week period of response collection, the data was exported as a CSV file.
We then wrote Python scripts for data cleaning and analysis.
To avoid too fast responses where participants could not have really grasped the task and too slow responses where a task interruption was very likely, we excluded all individual task responses that took less than 10 seconds or more than 400 seconds.
In total, four individual task responses were removed.
Additionally, the responses of one participant were discarded completely because the 400 seconds threshold was reached for 3 of 12 tasks, with a total participation time of 37 min, while the median was 14.5 min and the mean was 15.5 min.
For these \textbf{65 valid responses}, we then had to verify the correctness of the free-text answers.
The script provided an automatic check for the most common correct and incorrect answers, but the remaining responses for these tasks were graded manually.
Obvious spelling mistakes such as \textit{Main Srvice} instead of \textit{Main Service} were tolerated, but in the case of tasks with HTTP requests or source code, the corrections were very strict regarding the syntax.

For the analysis, we calculated TAU for all tasks, as well as the most common descriptive statistics.
The \textit{Shapiro-Wilk test}~\cite{Wilk} confirmed that 22 out of the 24 columns were not normally distributed.
We therefore selected the \textit{Mann–Whitney U test}, a non-parametric test for values of two independent groups~\cite{Whitney}.
Since we conducted six hypothesis tests (one per pattern), we used the \textit{Holm-Bonferroni correction}~\cite{Holm-Bonferroni} to combat the multiple comparisons problem~\cite{MultipleProblem}.
The reported p-values in the paper are the adjusted values.
We report effect sizes using \textit{Cohen's d}~\cite{Cohen}.
The interpretation of the values is based on \citet{Sawilowsky}:
\begin{itemize}
    \item $d \le 0.2$: very small effect
    \item $0.2 \le d < 0.5$: \text{small effect}
    \item $0.5 \le d < 0.8$: \text{medium effect}
    \item $0.8 \le d < 1.2$: \text{large effect}
    \item $1.2 \le d < 2.0$: \text{very large effect}
    \item $d \ge 2.0$: \text{huge effect}
\end{itemize}

For analyzing the impact of demographic attributes (RQ2), we first normalized the respective data and used bucketing in some cases, e.g., for years of experience.
We then used the fairly robust \textit{Kendall's $\tau$}~\cite{Kendall} to analyze correlations between demographic attributes and TAU.
Lastly, we used linear regression models~\cite{LineareRegression} via the \texttt{statsmodels} package\footnote{\url{https://www.statsmodels.org/stable/regression.html}} to predict TAU, which provided a more in-depth analysis of several attributes at once.

\begin{figure*}[ht]
  \centering
  \includegraphics[width=\textwidth]{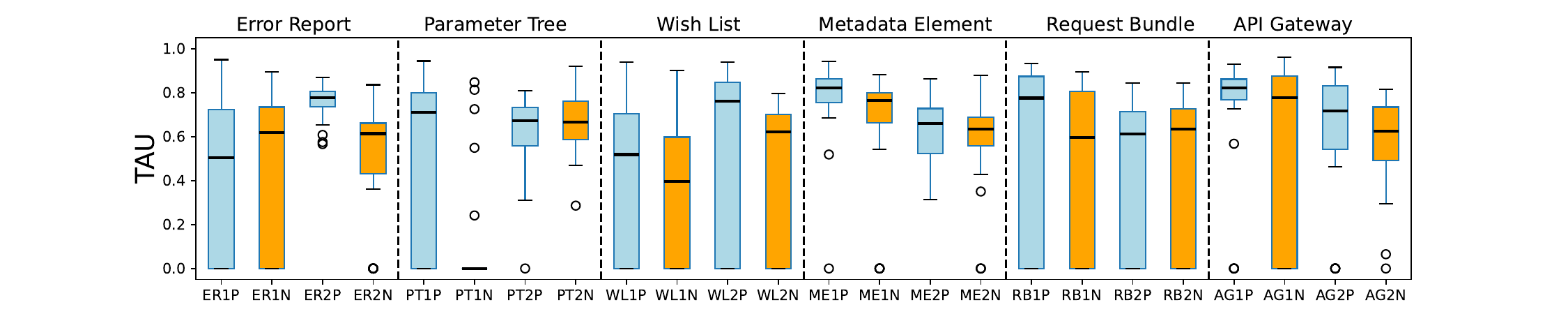}
  \vspace{-15pt}
  \caption{Boxplots of TAU values for all tasks; pattern versions \texttt{P} in blue (dark-gray in b/w print), non-pattern versions \texttt{N} in orange (light-gray in b/w print)}
  \label{fig:TAU-boxplots}
\end{figure*}

\begin{table*}[bp]
    \caption{Descriptive statistics for all 12 experiment tasks and versions (\texttt{P} and \texttt{N})}
    \label{tab:StatOverview}
    \centering
    \begin{tabular}{lrrrrrr}
        \multirow{2}{*}{\textbf{Task Name}} &
        \multicolumn{2}{c}{\textbf{Mean TAU}} &
        \multicolumn{2}{c}{\textbf{Correct Answers in \%}} &
        \multicolumn{2}{c}{\textbf{Mean Duration in s}} \\
        & Pattern (\texttt{P}) & Non-Pattern (\texttt{N}) & Pattern (\texttt{P}) & Non-Pattern (\texttt{N}) & Pattern (\texttt{P}) & Non-Pattern (\texttt{N}) \\
        \hline
        \hline
        Error Report 1 & 0.380 & 0.426 & 57.1 & 60.0 & 43.9 & 39.6 \\
        Error Report 2 & 0.759 & 0.500 & 100.0 & 82.9 & 41.5 & 67.5 \\
        Parameter Tree 1 & 0.510 & 0.091 & 66.7 & 17.1 & 38.2 & 56.7 \\
        Parameter Tree 2 & 0.626 & 0.672 & 100.0 & 100.0 & 38.6 & 33.9 \\
        Wish List 1 & 0.377 & 0.324 & 54.6 & 53.3 & 131.2 & 163.0 \\
        Wish List 2 & 0.548 & 0.387 & 70.0 & 60.0 & 80.9 & 119.1 \\
        Metadata Element 1 & 0.784 & 0.646 & 96.7 & 88.6 & 62.1 & 88.1 \\
        Metadata Element 2 & 0.639 & 0.557 & 100.0 & 93.3 & 53.4 &  54.1 \\
        Request Bundle 1 & 0.543 & 0.415 & 66.7 & 54.3 & 64.9 & 91.6 \\
        Request Bundle 2 & 0.459 & 0.427 & 74.3 & 63.3  & 45.5 & 39.3 \\
        API Gateway 1 & 0.729 & 0.586 & 91.4  & 72.4  & 84.3 & 76.9 \\
        API Gateway 2 & 0.605 & 0.584 & 80.0 & 97.1  & 42.8 & 55.9 \\
        \hline
        \hline
    \end{tabular}
\end{table*}

\section{Results}
As desired, our 65 participants had different backgrounds and levels of experience.
The reported professional experience ranged from 0 to 35 years, with a median of 4 and a mean of 6.6 years.
Overall, 46\% had 5 or more years of experience, i.e., our sample was quite experienced.
Regarding their professional background, similar shares of participants reported to be from academia and industry (40\% vs. 38.5\%), with the remaining 21.5\% answering to be active in both.
As their main role, 43\% declared to be students, with other prominent roles being software engineer (27.7\%), researcher (9.2\%), or software architect (7.7\%).
Experience in the three categories \textit{software design patterns}, \textit{HTTP APIs}, and \textit{MAPs} was recorded with a 5-point ordinal scale with the options \textit{novice (1)}, \textit{beginner (2)}, \textit{intermediate (3)}, \textit{advanced (4)}, and \textit{expert (5)}.
A bit more than a third of participants reported to be advanced or expert users of software design patterns (37\%) and HTTP APIs (35\%), with a median of 3 (intermediate) for both categories.
As expected, MAP experience was lower, with only 19\% reporting to be advanced and 3\% experts.
The median rating was 2 here (beginner).
When comparing these demographics between the two sequences (see Table~\ref{tab:demographics}),
we see that the randomization led to five more participants in sequence 1.
Additionally, sequence 1 had more students (17 vs. 11), fewer industry participants (8 vs. 17), but was overall slightly more experienced in all categories.
However, these differences were still fairly small, with the medians for the experience ratings being the same (3, 3, and 2).
Moreover, the nature of the crossover design ensured that each participant worked on both versions of a pattern task.

\subsection{Impact of Patterns (RQ1)}

\begin{tcolorbox}[left=2pt,right=2pt,top=2pt,bottom=2pt]
    \textbf{Findings for RQ1:}
    Five of the six patterns had a significant positive impact on understandability (exception:  \textit{Request Bundle}).
    Effect sizes were \textit{medium} for \textit{Parameter Tree} and \textit{small} for the other patterns.
\end{tcolorbox}

To quickly compare the experiment performance between the different versions \texttt{P} and \texttt{N}, we created boxplots per task that visualize the distribution of TAU (see Fig.~\ref{fig:TAU-boxplots}).
While this helps to quickly judge which version performed better, it is more difficult to derive a potential explanation from.
We therefore also provide the descriptive statistics for TAU, correctness, and duration per task and version in Table~\ref{tab:StatOverview}.
All in all, mean TAU was higher for the pattern versions \texttt{P} for 10 of the 12 tasks, the exceptions being \textit{Error Report 1} (ER1) and \textit{Parameter Tree 2} (PT2), even though there is also one more case with median TAU being lower for \texttt{P}, namely \textit{Request Bundle 2} (RB2).
In the following, we describe the results per pattern tasks in more detail.

Regarding the two \textit{Error Report} tasks, ER1 showed a similarly spread-out TAU distribution for both versions, but with visibly different medians.
As mentioned, this was one of the tasks where version \texttt{N} performed better, with answering both slightly more correctly (57\% vs. 60\%) and faster (44 s vs. 40 s).
For ER2, however, we see a clearly better performance for version \texttt{P}, with 100\% correct answers and considerably less required time.

For \textit{Parameter Tree}, PT1 displayed the most substantial difference between the two versions in all 12 tasks: \texttt{P} performed much better than the non-pattern version \texttt{N}, which displayed by far the worst performance.
Only 17\% answered correctly, which led to a mean TAU of 0.091.
PT2, on the other hand, was the second case where the non-pattern version \texttt{N} performed better, even though it was only a slight difference.
All participants answered correctly in both versions, but version \texttt{P} was a bit slower (39 s vs. 34 s).

In both \textit{Wish List} tasks, the pattern versions \texttt{P} performed better.
Correctness was similar for WL1 (55\% vs. 53\%), but \texttt{P} was faster (131 s vs. 163 s).
Overall, this task took participants the longest.
The performance differences were more pronounced in WL2, both for correctness (70\% vs. 60\%) and duration (81 s vs. 119 s).

We see a similar picture for both \textit{Metadata Element} tasks: the pattern versions \texttt{P} outperformed \texttt{N}, even though both versions had good results.
While both correctness and duration were better for \texttt{P} in ME1, the difference in ME2 mostly lies in the correctness (100\% vs. 93 \%), with duration being very similar (53 s vs. 54 s).

The two \textit{Request Bundle} tasks showed some differences, even though the mean TAU was slightly better for \texttt{P} in both cases.
In RB1, both correctness (67\% vs. 54\%) and duration (65 s vs. 92 s) were clearly favorable for \texttt{P}, but RB2 was more complicated.
While \texttt{P} had higher correctness (74\% vs. 63\%), they also required more time (46 s vs. 39 s).
Due to the substantial spread of TAU in both versions, this in the end even led to a slightly higher median TAU for \texttt{N}.

Lastly, the pattern versions \texttt{P} of the \textit{API Gateway} tasks performed slightly better, but due to different reasons.
In AG1, \texttt{P} answered correctly more often (91\% vs. 72\%), but needed slightly more time (84 s vs. 77 s).
TAU was also visibly more spread out for \texttt{N}.
However, in AG2, it was the other way around: \texttt{N} had better correctness (80\% vs. 97\%), but \texttt{P} was overall faster to compensate (43 s vs. 56 s).
In both tasks, these differences still led to a better TAU for the pattern version.

\begin{table}[ht]
    \caption{Mann–Whitney U hypothesis testing results (Holm-Bonferroni adjusted p-values, $\alpha=0.05$, ordered by effect size)}
    \label{tab:StatResults}
    \centering
    \begin{tabular}{lrrrr}
        \textbf{MAP} & \textbf{U value} & \textbf{p-value} & \textbf{Cohen's d} & \textbf{H$_{1}$ Accepted} \\
        \hline
        \hline
        Parameter Tree & 2823.0 & 0.0023 & 0.66 & Yes \\
        Metadata Element & 2531.5 & 0.0336 & 0.42 & Yes \\
        API Gateway & 2655.5 & 0.0167 & 0.29 & Yes \\
        Wish List & 2528.0 & 0.0336 & 0.28 & Yes \\
        Error Report & 2621.5 & 0.0336 & 0.27 & Yes \\
        Request Bundle & 2371.0 & 0.1092 & -- & No \\
        \hline
        \hline
    \end{tabular}
\end{table}

Despite the better mean and median TAU for the pattern version \texttt{P} in most of the 12 tasks, a combined hypothesis test per pattern was still necessary to identify if these differences really were statistically significant.
We present the results of the Mann–Whitney U tests in Table~\ref{tab:StatResults}.
Since we used the Holm-Bonferroni correction to adjust the p-values, the required significance level $\alpha$ is 0.05.
We also provide Cohen's $d$ for the effect size.
Based on the test results, we accepted the alternative hypothesis H$_{1}$ in five cases, i.e., \textbf{we found a significant positive impact on understandability for five of the six examined patterns}.
The only exception was \textit{Request Bundle}, where the adjusted p-value of 0.1092 was too large to reject the null hypothesis.
Regarding effect sizes, \textit{Parameter Tree} showed a \textit{medium} effect (0.5 $\le d <$ 0.8), with the other four patterns leading to a \textit{small} effect (0.2 $\le d <$ 0.5).

\subsection{Influence of Demographic Data (RQ2)}

\begin{tcolorbox}[left=2pt,right=2pt,top=2pt,bottom=2pt]
    \textbf{Findings for RQ2:}
    Years of experience was not correlated with experiment performance.
    Being a student was negatively linked to performance in \texttt{P} but not in \texttt{N}.
    Best predictors for performance were MAP experience (stronger in \texttt{P}), general pattern experience (stronger in \texttt{N}), and HTTP API experience (stronger in \texttt{N}).
\end{tcolorbox}

For the exploratory RQ2, we wanted to identify which demographics influenced the effectiveness of the studied patterns.
We therefore calculated the mean TAU per participant separately for the two versions \texttt{P} and \texttt{N}, and then calculated a correlation matrix with the following attributes: years of experience, being a student, being from industry, general pattern experience, HTTP API experience, and MAP experience.

Interestingly, there was no significant correlation between years of experience and TAU for both versions, i.e., having worked for longer was not linked to better performance.
For \texttt{P}, being a student was negatively correlated with TAU ($\tau=-0.236$, $p=0.0219$), meaning students were linked to slightly worse performance in the pattern tasks.
This was significant at $\alpha=0.05$, but would not hold up to more rigorous p-value adjustments.
No such significant correlation existed for the non-pattern tasks \texttt{N}.
Conversely, being from industry was positively correlated with TAU in both versions
(\texttt{P}: $\tau=0.277$ and $p=0.0070$, \texttt{N}: $\tau=0.227$ and $p=0.0270$).
Both were significant at $\alpha=0.05$, but the correlation in \texttt{P} was slightly stronger, i.e., it helped more to be from industry for the pattern tasks.
However, TAU showed the strongest correlations in both versions with the self-reported experience categories, all of them significant.
MAP experience was unsurprisingly more important for \texttt{P} ($\tau=0.294$, $p=0.0019$) than for \texttt{N} ($\tau=0.249$, $p=0.0084$).
General pattern experience, on the other hand, showed a stronger correlation in \texttt{N} ($\tau=0.427$, $p<0.0001$), even though it was also important in \texttt{P} ($\tau=0.378$, $p<0.0001$).
Lastly, experience with HTTP APIs was the strongest individual predictor for better performance.
This link, again, was slightly stronger in N ($\tau=0.435$, $p<0.0001$) in comparison to \texttt{P} ($\tau=0.403$, $p<0.0001$).
To analyze the impact of all of these predictors simultaneously, we built two linear regression models predicting TAU, one for \texttt{P} and one for \texttt{N}.
Both models were able to explain a decent amount of the variance.
However, the model for \texttt{P} showed an adjusted $R^2=0.334$, while it was $0.444$ for \texttt{N}, indicating that the model for tasks without patterns was able to explain substantially more variance in the experiment performance based on demographics.

\section{Discussion}

We provided support that five of the six examined MAPs had a significant, albeit small to medium effect on understandability.
The exception was \textit{Request Bundle}, where the pattern version \texttt{P} was slightly better in the mean, but not in the median for \textit{Request Bundle 2} (RB2), ultimately leading to an overall non-significant test result.
Additionally, the non-pattern version \texttt{N} also performed slightly better for two individual tasks, namely \textit{Error Report 1} (ER1) and \textit{Parameter Tree 2} (PT2).
This also indicates that the degree of effectiveness of a pattern is very much related to the concrete use case and implementation.
However, these differences were small enough and the performance of \texttt{P} in the complementary tasks ER2 and PT1 was so much better that the combined tests still confirmed significant differences.
One possible reason for the better performance of \texttt{N} in these instances might be the slight overhead in complexity that each pattern introduces.
This could manifest in the form of additional required time to understand the respective pattern for the first time.
Such an effect would gradually subside with the pattern becoming familiar, which was irrelevant in our experiment due to the singular exposure to each pattern.
For example, in the case of RB2, correctness of \texttt{P} was more than 10 percentage points higher, but mean duration also increased by 6 seconds compared to \texttt{N}.
In addition to the added complexity, using copy \& paste probably alleviated some effort of creating four simple API requests for \texttt{N} instead of one more complex one for \texttt{P} (see Fig.~\ref{fig:FreeTextTask}).
For PT2, correctness was similarly both at 100\%, but \texttt{P} needed on average roughly 5 seconds more, adding further support to the complexity hypothesis.
We theorize that a significant impact of \textit{Request Bundle} and stronger effect sizes for the other patterns would be visible after repeated exposure and familiarization, which would typically happen with continuous involvement in the development of a software system.
Since every pattern version only appeared once per participant, we lack the necessary data to support this, i.e., additional longitudinal research is required to confirm this.

\begin{tcolorbox}[left=2pt,right=2pt,top=2pt,bottom=2pt]
    \textbf{Implication:}
    MAPs can introduce complexity that increases the time required for understanding, even though the gained understanding can be more correct.
\end{tcolorbox}

This experiment took place in a controlled, simplified environment that is clearly different from a realistic setting in which software engineers use or integrate an API.
Such a real-world environment would have many more influencing factors, a different way to present the APIs, and also different ways to interact with the APIs.
Understandability and API usability also would not be the only quality concerns that would be evaluated.
Patterns typically are concerned with trade-offs, i.e., they can improve some quality attributes while negatively impacting others.
For example, \textit{Request Bundle} clearly improves performance efficiency by reducing the number of API requests~\cite{ElMalki2021}, even though we found no evidence for a positive impact on understandability.
Moreover, real-world APIs often do not contain a single pattern in isolation, but can incorporate multiple patterns, adding to the complexity of their effects.
Nonetheless, we see a strong need to analyze the different MAPs in a controlled environment to provide evidence for their effectiveness.
If they do not show their potential to improve understandability and API usability under such conditions, their usage, or necessary preconditions thereof, need to be questioned, as is the case for \textit{Request Bundle}, which requires further research.
But even for the four patterns that showed significant impact, the fairly small effect sizes under such favorable conditions should give rise to further research to identify clear requirements for them to be effective.

\begin{tcolorbox}[left=2pt,right=2pt,top=2pt,bottom=2pt]
    \textbf{Implication:}
    In real-world environments, MAPs are concerned with trade-offs, e.g., while \textit{Request Bundle} did not improve understandability, it can improve performance.
\end{tcolorbox}

Regarding the influence of demographics on pattern effectiveness (RQ2), we identified no link with years of experience, but being a student was linked to worse performance in \texttt{P}.
The link between being from industry and better performance existed in both versions, but was stronger in \texttt{P}, with the same being true for MAP experience.
All of this seems to support the complexity hypothesis: when confronted with an API example that contains a pattern (\texttt{P}), a person experienced with this or a similar pattern will profit more from its effects on understandability and API usability.
However, when such a pattern is absent (\texttt{N}), other factors become more helpful to understand and use the API, as evidenced by the stronger links for general pattern and HTTP API experience in \texttt{N}.
The importance of experience with HTTP APIs for experiment performance also underlines the impact of the chosen technologies to implement a pattern.
Finally, the better explanatory power of our performance prediction model for \texttt{N} (11 percentage points more compared to \texttt{P}) might further indicate the complexity of influences and variety of required experiences when a pattern is present, making its effect difficult to predict.
It could be that experience variables for individual MAPs (which we did not collect) might have improved the prediction quality for \texttt{P}.
All in all, these results highlight the importance of training and education about MAPs, both in industry and academia, to fully leverage their beneficial effects, which has clear implications for teachers and for the onboarding of new engineers.

\begin{tcolorbox}[left=2pt,right=2pt,top=2pt,bottom=2pt]
    \textbf{Implication:}
    Education and documentation of MAPs are prerequisites for their effectiveness in industry.
\end{tcolorbox}

\section{Threats to Validity}
Even though we carefully constructed and refined our experiment design after a pilot with API experts, there are still some threats to validity in different areas that we need to mention.

\textit{Construct validity} is concerned with the alignment of the studied concepts and the operationalized experiment measures, i.e., did we really measure what we intended to measure?
We wanted to analyze the impact on understandability, which extended towards API usability in some tasks.
There, it was not enough to grasp the purpose of the API example (the learnability component of usability), but participants also needed to construct statements to interact with the API.
The primary used measure, namely Timed Actual Understandability (TAU) as a combination of correctness and time~\cite{DefinitionTAU}, is an accepted metric for controlled experiments that has been applied in several comprehension experiments.
TAU leads to higher information density per individual hypothesis test, but its trade-off between time and correctness also leads to unusual distributions for binary correctness.
These limitations can be mitigated by separately investigating correctness and time during results interpretation.
Moreover, other concepts that are common in the space of microservices, APIs, or HTTP requests were not explicitly added to the tasks for the sake of simplicity and control.
At the start of the experiment, participants were informed that the goal of this research was to examine the impact of some API design patterns on software quality, with the term \enquote{MAPs} or specific pattern names not being mentioned.
While this definitely left some probability for \textit{hypothesis guessing}~\cite{Threat:Construct}, we believe it to be very unlikely that some participants identified their respective task versions and then adjusted their answers accordingly.

\textit{Internal validity} can be impacted by confounding factors that secretly influence the dependent variables.
Because it limits the influence of inter-participant differences, e.g., regarding prior knowledge or motivation, our crossover design offered a decent degree of robustness against several typical confounders.
Our experiment had a mean duration of 15.5 minutes and should therefore have restricted the number of participants that prematurely stopped because they got bored, fatigued, or demotivated, i.e., \textit{mortality}~\cite{Wohlin2012}.
Additionally, our experiment included some facets of randomization, e.g., the assignment of participants to one of two task sequences and the order of possible answers in single-choice tasks.
The two task sequences were consciously arranged in a way to minimize potential threats regarding \textit{familiarization}.
Learning effects are still very likely because the task structures were fairly similar.
However, harmful carryover effects are still not expected due to the static sequences that ensured that both versions of the same task received the same position.
This guaranteed a similar maturation per pattern task.
Additionally, we put the maximum distance between the two tasks related to the same pattern, e.g., ER1P and ER2N, thereby allowing the maximum amount of time to defamiliarize.
While it is possible that some participants may have seen some familiarity with the first task, the different API snippet made it unlikely that such recognition distorted the results.
Despite the randomized seeding of participants, the two sequences still showed a minor imbalance regarding experience, which could have impacted individual pattern tasks.
The crossover design with two tasks per pattern should have reduced this effect, though.
Furthermore, the absence of \textit{random irrelevancies}~\cite{Wohlin2012} cannot be guaranteed because the experiment was conducted online, with no control over the participants' environment.
To minimize this threat, we used the required time to discard participant responses considerably longer than the average.
Overall, this increased variance does not seem to have influenced our hypothesis testing.
The reduced control and observation in online experiments also means that participants could easily cheat, e.g., use the help of others or participate multiple times.
However, online experiments are accepted practice in SE research because there are typically no incentives for cheating.
We therefore perceive threats in this area as negligible.

Lastly, \textit{external validity} is concerned with the generalization potential of the results, i.e., if they are transferable to other settings or population samples.
To increase the diversity and number of participants, we conducted an online experiment via a survey tool.
Additionally, the variety of participants was increased by including people from both academia and industry, as well as MSc and PhD students.
Our tasks covered six microservice patterns from the perspective of API usability and understandability.
A generalization to other software design patterns or quality attributes is not feasible.
With 65 valid responses and the crossover design, our sample size was satisfactory.
However, more participants could have further increased the generalization potential of the results.
Finally, our tasks were all formulated with great care and double-checked by API experts so that they show a decent degree of realism.
Nonetheless, there are obvious differences between the experiment environment and a real-world setting, which we also touched upon in the discussion section.
In practice, developers could look up pattern documentation and HTTP tutorials, or ask colleagues for help.
But even in the real world, such materials or documentation of the used patterns may not be present.
Consulting documentation or a colleague also requires more time, so it would still be better to comprehend an API without this.
In general, we think that our results offer a decent degree of realism to allow comparisons with less controlled industry environments.
Follow-up research is necessary to confirm the full extent of this generalization.

\section{Conclusion}
To provide empirical evidence for the impact of microservices patterns on understandability, we conducted a controlled experiment with 65 participants.
For five of the six studied patterns, we found a significant positive impact, with effect sizes ranging from small to medium.
While we found significant effects for most MAPs that encourage their usage in industry, our results also highlight the importance of familiarity with the patterns to combat their slightly increased complexity.
This requires educational efforts in academia, but should also inform the onboarding of new hires in practitioner teams responsible for systems containing MAPs.
A clear documentation of the used patterns and knowledge sharing around them are prerequisites for this.
Moreover, longitudinal research with MAPs should provide further support for the complexity and familiarization hypothesis.

Since one controlled experiment can only study so much, future research needs to provide empirical evidence for other MAPs regarding various important quality attributes, e.g., usability of MAPs not covered by us and maintainability or reliability of MAPs, which seems to be an understudied area.
Combinations of patterns from a single language (such as MAPs) and patterns from different languages, as well as the influence of experience with a concrete pattern, but also trade-offs related to individual patterns are important areas, for which we currently do not have enough evidence.
Lastly, to strengthen the knowledge we have in this space, we encourage replications of this experiment with different samples of the population or adapted API examples.
To allow such replications, but also for transparency reasons, we share our experiment and analysis materials on Zenodo.\footnote{\url{https://doi.org/10.5281/zenodo.10210246}}.

\section*{Acknowledgment}
We kindly thank all six experts who participated in our pilot and provided feedback on the experiment design.
We also thank all our experiment participants for their time!
Lastly, we thank MAP author Mirko Stocker (Eastern Switzerland University of Applied Sciences) for reviewing a draft of this paper.

\bibliographystyle{IEEEtranN}
\bibliography{references}

\end{document}